# Is There a Wave Excitation in the Thalamus?

Robert Worden

UCL Theoretical Neurobiology Group

rpworden@me.com

September 2019

Abstract:

This paper proposes that the thalamus is the site of a wave excitation, whose function is to represent the locations of things around the animal. Neurons couple to the wave as transmitters and receivers. The wave acts as an analogue representation of local space. This has benefits over a purely neural representation of space.

Several lines of evidence support this hypothesis – both theoretical, concerning efficient Bayesian inference in the brain – and empirical, concerning the neuro-anatomy of the thalamus. Across all species, the most basic function of the brain is to coordinate movements in space. To represent positions in space only by neural firing rates would be complex and inefficient. It is possible that that the brain represents 3D space in a direct and natural way – by a 3D wave.

## 1. Introduction

This article introduces a radical proposal; namely, that there are physical waves in the thalamus that encode the three-dimensional geometry of our lived world. At first glance, this may seem an outlandish proposal, given our current understanding of functional brain architectures and their neurophysiology. However, a closer inspection of this 'Thalamic Wave Hypothesis' suggests that it should be taken seriously as a generalisation of established principles of functional anatomy – a generalisation that not only resolves some deep issues about the nature of how we represent space, but also enjoys a remarkable amount of empirical support.

The notion of place coding and subsequent representational maps in the brain is well established [Chen et al., 2014; Lisman and Buzsaki, 2008; Livingstone and Hubel, 1988; Moser et al., 2015; O'Keefe and Recce, 1993; Stachenfeld et al., 2017; Zeki, 2005; Zeki and Shipp, 1988]. The most celebrated example of this is the retinotopic mapping that defines the functional architecture of early visual cortex [Maunsell and Treue, 2006; Nealey and Maunsell, 1994; Zeki and Shipp, 1988]. The Thalamic Wave Hypothesis takes this idea further – to the encoding of three-dimensional spatial information. I will focus on the spatial encoding and geometry that is a necessary part of our internal or generative models [Rudrauf et al., 2017; Williford et al., 2018].

If any computing mechanism obeys the same physical laws as the subjects of its processing (its domain of the computation), there can be major benefits in computational efficiency. The results of the computation are physically constrained to obey constraints of the domain. Appropriate computations then result from the underlying physical mechanism, and do not need to be designed into the computing architecture. This can dramatically reduce the complexity of the design space for computers, while also furnishing more efficient computation.

This paper suggests that when the brain computes about 3D local space, it uses an analogue representation of 3D spatial positions – stored in a 3D wave.

Because of their advantages, analogue computers were used until the 1960s – after which they were eclipsed by the remarkable success of digital computers, resulting from Moore's law [Moore 1965]. There is no Moore's law for animal brains. While neurons act as a universal computing engine in brains, they have costs, such as brain energy costs which are significant for many animals. We should not discount the idea that some aspects of animal cognition use analogue computing mechanisms [Sloman 1975].

In what follows, I hope to unpack the advantages of analogue 3D spatial computation from first principles. A related account of how nervous systems (self) organise to become a living model of their world can be found in early accounts of cybernetics; e.g., the good regulator theorem [Conant and Ashby, 1970; Seth, 2015], right through to current formulations of the Bayesian brain in terms of concepts like the free energy principle [Friston et al., 2006]. The underlying theme here is that to survive and navigate in a world – given just sensory impressions of that world – it is necessary to have an internal model of the world. In turn, this implies that certain aspects of the 'model' (i.e., brain) recapitulate the causal structure behind the sense data. A basic foundation of that structure is three-dimensional Euclidean geometry.

The proposal here is that there are (non-neuronal) dynamics in the brain (i.e. the thalamus) that maintain a one-to-one mapping with the geometry of the world causing our sensations. In what follows, I rehearse the evolutionary imperatives for this sort of representation from both a biological and computational perspective. I will then review the anatomical and physiological evidence in support of the Thalamic Wave Hypothesis before turning to its explanatory scope – in terms of comparative anatomy, evolution and the form of consciousness.

Waves can serve as an analogue representation of positions, in that wave vectors obey the same 3D geometric constraints as positions in space. Components of the wave and points in space are related by a Fourier transform, which preserves geometric relations (as in a hologram). In representing 3D positions of objects, a wave may have major advantages over a representation in neural firing rates:

- Simplicity of design
- High capacity, to represent positions of many objects (or the probability density over many positions of one object)
- High spatial precision
- Geometric fidelity
- High input-output connectivity
- High speed of storage and retrieval
- Direct computation of geometric displacements (3-D vector subtractions)

There are several lines of empirical evidence for a wave representing positions in the thalamus. These are unpacked in the later sections of the paper:

- **Neural connectivity of the thalamus**: the thalamus is a relay station for just those sense data required to fix the positions of objects; but it does not relay sense data not used for object location (notably olfaction)
- **Shaped of the thalamus and TRN**: Across many species, the thalamus has approximately equal spatial extension in all three dimensions – suitable for holding a wave – and the thalamic reticular nucleus (TRN) has the form of a thin shell around the thalamus (suitable to be a transmitter of the wave, immersing the volume of the thalamus)
- **The anatomy of the thalamus and the TRN only make sense in a wave hypothesis**: The thalamus has many nuclei, with no direct connections between them. Why are they close together? Axonal energy costs would be reduced (over evolutionary time) if thalamic nuclei migrated out towards cortex – if the thalamus exploded. Similarly, the shell-like form of the TRN is uneconomical in axonal connection lengths. This anatomy (found across many species) only makes sense if all the thalamic nuclei need to be immersed in the same wave.
- **The form of consciousness**: A large part of consciousness takes the form of a faithful 3D geometric model of one's present surroundings. If consciousness arises from the thalamic wave, this gives a simple and testable account of the core experimental fact about consciousness – its 3D geometric nature.

If the wave hypothesis is correct, it has profound implications for neuroscience. It would imply that many of the most important functions of any animal brain – any function involving spatial cognition – are accomplished not by neurons alone, but by neurons coupled to the wave.

## 2. Spatial Cognition – the Requirement

Since animals first had limbs and complex sense data – in the Cambrian era, some 500 million years ago – brains have been subject to some simple, universal, and highly demanding requirements, to make use of the sense data and control limbs.

An imperative for the animal brain is to control muscular movements. This involves knowing where things are, and what they are, in 3D space. The things include your own limbs, so you can make appropriate actions towards the other things you perceive – move towards them, or move around them, or bite them, or grasp them, or strike them.

These are the core things that every animal brain needs to control, at every moment of the day. The selection pressure to do them right is very strong indeed. Yet the requirements to do it well are very stringent:

- Things are located in three dimensional space: so the brain needs to represent locations in 3D space – with three degrees of freedom for every represented point.
- Locations obey the constraints of 3D geometry: for instance, if the displacement from object i to object j is the 3-vector $\mathbf{a}_{ij}$, then $\mathbf{a}_{ik} = \mathbf{a}_{ij} + \mathbf{a}_{jk}$, for any i, j, and k.
- Most of the time, most things do not move. This core (Bayesian) prior determines which things need paying attention to – the things that are really moving in 3D allocentric space. It also economizes on sense data – if something is not moving, you need not keep checking where it is. The constraint does not apply to raw sense data (which change for many reasons) only to a model of objects in 3D space which can be inferred from it[1].
- Things around the animal need to be classified, into classes which matter to the animal – food, potential mates, predators, shelter, and so on.
- The things in any class have similar 3D shapes, wherever they are relative to the animal. This Bayesian prior is enormously useful both in learning to recognize those shapes (allowing learning to be orders of magnitude faster, using far fewer learning examples) and then in recognising the shapes from moment to moment. Again, this is not a constraint on raw sense data; it is a constraint on a 3D model inferred from sense data.

The last requirement can be stated as a requirement for **translationally invariant** pattern learning and pattern recognition – to learn and recognize the same classes of thing, wherever they are in 3D space around the animal. It implies a requirement for **spatial steering,** to steer sense data arising from anywhere around the animal, to modular learning and recognition units in the brain [Olshausen et al, 1993, 1995]. This signal steering should be free of place-dependent distortion.

These universal constraints have applied to all vertebrate brains since the Cambrian era. They have exerted enormous cumulative selection pressures on the design of brains. We would expect brains to have adapted very well to them. We know that very effective brain designs to have evolved - so brains have adapted well to these requirements.

Yet in terms of neural computing architecture, the requirements are far from trivial. It is not obvious how a neural computer might embody these constraints. To cite one measure of this difficulty: while modern neuroscience and AI have made great strides in understanding some aspects of cognition (such as language) the progress in spatial cognition has at best been modest. The leading examples of the art – convolutional neural nets [LeCun, Bengio & Hinton, 2015] – employ algorithms which have little biological relevance. They use huge computing power, and require huge and biologically infeasible numbers of training examples.

---

[1] Technically, this means that there is a minimisation of complexity that is inherent in Bayesian inference (i.e., log Bayesian evidence is accuracy minus complexity). Having a low dimensional representation of allocentric space therefore provides the most efficient, nonredundant or minimally complex inference.

The essence of the five requirements above is to represent the 3D locations of objects in space, in a way that obeys the constraints of 3D Euclidean geometry. Besides using neural firing rates to represent 3D positions (which is the mainstream hypothesis of modern neuroscience), there is an alternative approach to meeting those requirements.

### 3. The Wave Hypothesis

This section sets out the core hypothesis of the paper - stated for mammals, although its application may be wider. Letters in bold type represent 3-vectors. The hypothesis has the following components:

- There is a wave excitation in the thalamus (whose physical nature is not yet known)
- The wave is a superposition of many components with different wave vectors **k**.
- A component with wave vector **k** represents an object at position **r** = λ**k** in some allocentric frame of reference (the frame is chosen so that in the short term most objects do not move, most of the time)
- Neurons couple to the wave, as transmitters and receivers
- The Thalamic Reticular Nucleus (TRN), surrounding the dorsal thalamus, acts as a transmitter for the wave, so that the wave immerses all dorsal thalamic nuclei
- Neurons in dorsal thalamic nuclei (particularly in secondary nuclei) act as receptors for the wave, sending outputs to cortex.
- Individual thalamic receptor neurons may be sensitive to a range of wave vectors **k**
- Neurons may be tunable, allowing their range of **k** to change dynamically.

These points are shown in a diagram:

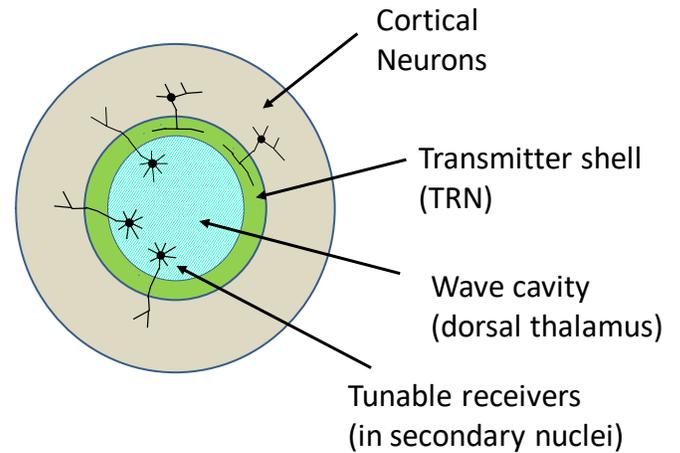

While the physical nature of the wave is not yet known, there are many kinds of quantized excitation known in condensed-matter physics, and they have been explored in biological material [Frohlich 1968, Davidov 1977, 1985]. It would be premature to suppose, just because we have not found an obvious candidate for the thalamic wave, that it could not have evolved.

Furthermore, we would expect the wave to exist now at low intensities, having evolved to economise on brain energy consumption [Attwell & Laughlin, 2001; Chklovskii et al 2002] ; we should not expect to detect it easily, without a clear idea of what we are looking for.

One example of a non-synaptic interaction between neurons is ephaptic coupling [Jefferies 1995] which involves local electric fields.

### 4. Computational Benefits of Wave Storage of Spatial Data

To compare a wave storage of 3D positions with a purely neural storage mechanism, we need some proposals for what the neural mechanism might be. This presents some difficulty, as there has been little theoretical work to define a neural 3D storage mechanism, or empirical work to test it.

There are many two-dimensional neural maps in the brain, [Livingstone & Hubel 1988; Nealey & Maunsell 1994; Maunsell & Treue 2006; Zeki 2005] where the location of a neuron in some sheet denotes a 2D position. This is an analogue representation of 2D position, and we might try to extend it to three dimensions, in one of two possible ways:

- Position is denoted by the location of a neuron in some 3D cluster of neurons.
- Two dimensions of position are denoted by the position of a neuron in a sheet; and the third dimension (e.g. depth) is defined in some other way.

We can perhaps discount the first option because:

- a 3D cluster of neurons would pose severe difficulties of connectivity for neurons near the centre of the cluster.
- No such cluster of neurons has been observed anatomically or physiologically in the form of organised, map-like[2] 3D 'place cells'.
- Such a 3D 'place by place' representation would show obvious place-dependent lesion effects, which as far as I know have not been observed.

We are left with two options:

a. Some highly distributed architecture, in which information about the location of one object in space is distributed across large numbers of neurons.
b. A representation in which at least one dimension of the position of an object is represented by the firing rates of some small set of neurons - as in a '2D map + depth' representation, or a 'triplet' representation in which all three dimensions of a position are represented by firing rates.

A fully distributed architecture (a) has interesting potential, but has hardly been investigated. I therefore focus on possible neural representations (b) in which one or more dimensions of position are represented by firing rates. Besides the problem of choice of a coordinate system, these neural representations face major challenges:

1. **Spatial precision versus speed**: In the simple case where some dimension of an object's position is represented by a single neuron's rate of firing, over the time period where that neuron emits 100 spikes (say of the order of 1 second), the statistical precision of the coordinate[3] is approximately one part in 10. Precision only increases as the square root of the number of spikes. For small mammals, this precision seems inadequate. There may of course be more complex representations of the coordinate, but those representations would compound the next challenge.
2. **Computing spatial displacements**: It is a core requirement that when the positions **a** and **b** of two objects are represented, some modules of the brain (such as those for pattern learning and recognition) should have rapid and precise access to the spatial displacement (**a**-**b**) between them [Olshausen et al. 2003]. Computing any coordinate of (**a**-**b**) requires precise subtraction of distances, particularly when **a** and **b** are close together (e.g. recognising some distant object; or computing its 3D shape from relative motion; or guiding limbs relative to a visual target). Any representation other than a simple 'single firing rate' representation makes this a complex computation.

These requirements, which together form a major challenge for any neural representation of positions, can be met by a wave representation:

a. **Spatial precision**: If the volume containing the wave has spatial extent D, and the minimum possible wavelength is λ, then the achievable spatial precision is approximately one part in (D/λ), which can easily be a large number. If a neuron within the wave region has receptors over a spatial region with smaller extent d, then that neuron can be tuned to a range of wave vectors (i.e. a range of represented positions), with spatial precision (d/λ), which may still be large.
b. **Speed**: Speed of storage and retrieval is bounded by the frequency of the wave, which is unknown, but may be high. After a few wave cycles, the wave can be set up and receptor neurons can detect it. There is no difficult speed/precision tradeoff.
c. **Computing Spatial Displacements**: To meet this requirement, it is simplest to have two kinds of wave, a 'signal' wave (which represents many objects at positions **k**) and an 'attention' wave which has a narrow range of wave vectors around some 'searchlight' wave vector **s**. Then if neurons are receptive to the interference between these two waves, Cos(**k.x**) + Cos(**s.x**) ~ Cos( **(k-s).x**). Thus the interference pattern simply computes the displaced position vectors (**k - s**), with no spatial distortion.

The 'heterodyne principle' in (c) directly calculates spatial displacements, and also gives a simple, distortion-free solution to the problem of spatial steering of signals – for example, steering the visual inputs from a face (wherever it may be relative to the animal) – to some face recognition module in cortex. Distortion-free spatial steering is essential for rapid learning of spatial patterns, and for recognising them once learnt.

The key benefits of wave storage of positions - storing multiple object positions with high precision and low spatial distortion - are directly demonstrated by holography.

One clear example of the requirement for precise storage of depth information, and calculation of small displacements in depth, is the requirement to compute **shape from motion**. [Murray et al. 2003] This is the use of visual data (without stereopsis), together with a strong Bayesian prior probability that most objects are rigid bodies, to compute the 3D shape of an object from the relative motion of its parts (as caused by motion of the object, or of the viewer).

---

[2] The 3D place cells observed in the bat hippocampus by [Wohlgemuth et al 2018] appear, like other place cells [O'Keefe and Recce 1993], not to be topographically organised as a map of local space; so those cells do not appear suitable to store a 3D geometric model of local space - unlike, say, the V1 visual cortex which does represent 2D geometry.

[3] Clearly, this argument about spike rate codes only applies to single neurons. It is possible that place codes could be encoded by populations of neurons, whose precision would increase with the number of neurons in each ensemble. For simple encodings, precision only increases as the square root of the number of neurons. Complex encodings make processing more complex.

This capability allows us rapidly to infer the 3D shape of a small, distant, rotating object, from its moving visual 2D projection. It can be modelled computationally as a Bayesian maximum likelihood fit to visual sense data, using a strong prior probability that the body is rigid. However, the actual computations, for a small or distant rotating body (e.g. occupying only a few degrees of the visual field), require high precision and speed in representing and computing the depth displacements of points on the body. If the precision is degraded (for instance by random neural noise in the depths), the computation rapidly falls apart, and can infer no shaped from motion.

In the example of shape from motion, wave storage of positions (including inferred depths) has obvious advantages of precision and computability, compared with the representation of positions in neural firing rates. The same advantages carry over to other types of spatial cognition; but they are particularly clear-cut when inferring shape from motion, for which no learning is required.

The analogue wave representation of 3D positions avoids the complex design choices of a pure neural architecture and has intrinsically high performance for several of the key requirements (spatial precision, fast representation; computing precise spatial displacements, low spatial distortion). The wave architecture to do these things is simpler and has better performance, than a purely neural architecture.

The basic reason for this is the huge intrinsic advantage of an analogue computer. Waves and spatial positions are related by a simple analogue relationship (a Fourier transform), so the geometry of the wave vectors naturally represents the geometry of real space. The required computations emerge naturally from the physics of the wave and transducers. They do not need to be designed into neural connectivity and firing rates.

There are other benefits. A wave representation can represent the positions of many objects at the same time, and has intrinsically high connectivity between its inputs and outputs. Any output neuron is connected to all input neurons, through the wave, but in a spatially selective way.

These core requirements for spatial cognition have been in place since the Cambrian Era. Even a small advantage would have favoured the wave storage architecture at many periods over that long timescale, and it could have evolved many times over. But if the advantage of a wave is large, more probably it evolved once, and has dominated since then.

An example of how geometry and waves get into neuronal architectures is the demonstration that lateral connections in the visual cortex are set up by travelling waves in the retina [Wong 1999]

## 5. A Hub and Spoke Architecture

This section summarises a related paper [Worden 2018].

Many cortex-centric models of cognition assume that there is peer-to-peer collaboration between different modular cortical knowledge sources. Different types of knowledge source include stereopsis, edge detection, shape recognition, shape from shading, multi-sensory integration and so on. There are probably may instances of each type of knowledge source (to apply in parallel to several objects around the animal), and there is evidence that different knowledge sources occupy different cortical regions.

Optimal Bayesian spatial cognition would combine the constraints from all these knowledge sources into a single maximum-likelihood fit to the 3D positions of all objects around the animal, integrating sense data of all modalities.

However, if this were done by peer-to-peer collaboration of all the knowledge sources, it would be inefficient in several ways:

- The absolute 3D position of each object would need to be stored redundantly in several different knowledge sources, with continual cross-updating of the best estimated position.
- To link N different knowledge sources in a peer-to-peer manner, there would need to be of the order of $N^2$ high bandwidth connections to convey the position messages between them.
- Even in the approximation of Gaussian errors (quadratic negative log likelihoods), finding the best fit to the 3D positions of N objects involves inverting a 3N by 3N matrix, which is an expensive computation.

These reasons imply that the peer-to-peer architecture is not the best one, and that a hub-and-spoke architecture (also known as star architecture or a master slave network) will perform better. With its extensive two-way connections to cortex, the thalamus is an obvious candidate for the hub, holding a single 'master' copy of the 3D location of each object or feature of interest, and acting as an 'aggregator' of constraints on the positions of things from all knowledge sources [Worden 2018]

This aggregator model of spatial cognition is shown in the diagram.

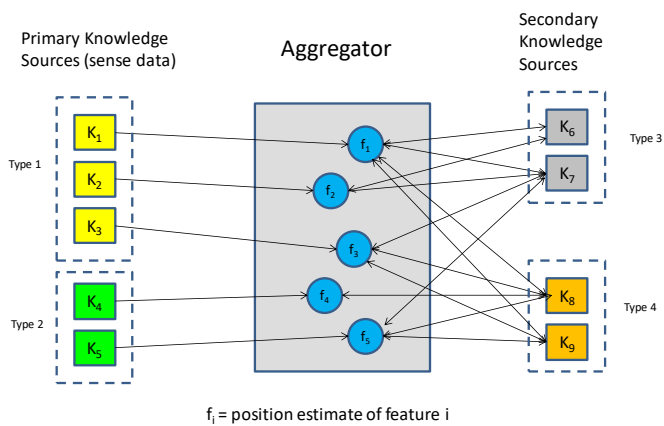

$f_i$ = position estimate of feature i

In this model, the many cortical knowledge sources do not need to communicate position information directly to one another (there is no $N^2$ communication problem), but only to the central aggregator. Each knowledge source can apply its own constraints on relative positions in an efficient., local process. The aggregator applies the global constraints to reconcile the constraints from different knowledge sources, separately for each feature of interest.

Computationally this is efficient, but it requires high spatial resolution and performance from the aggregator. As described above, this can be achieved using wave storage of positions in the thalamus.

Compared to the hub-and spoke architecture, the peer-to-peer cortical architecture is uneconomical not only in connections (and therefore brain energy consumption) but also in complexity – the number of different connections that need to be specified [epi]genetically in the design of the brain. By comparison, peer-to-peer architecture does not score well in simplicity of brain design.

Another advantage of the hub-and spoke architecture is that it offers a possible solution to the long-standing problem of feature [Treisman & Gelade 1980; Shadlen & Movshon 1999]: how do the different knowledge sources form temporary coalitions to find the best-fit positions of different features, and how is the computation organized so that different knowledge sources are dealing with the same feature at the same time? In this model, different features are bound by spatial position: if their estimated 3D positions overlap strongly in the aggregator, they are treated as parts of the same object.

## 6. Simplicity of Design and Evolution

It is notable that the wave storage of spatial positions can be described by the rather simple diagram of section 3, and by the simple mathematics of the Fourier transform. In contrast, a purely neural architecture for spatial cognition would involve many design choices, such as choices of coordinate systems, encoding of position coordinates, neural connectivity at many levels, and so on. A pure neural brain design appears to be much more complex than a design using wave storage for spatial cognition.

Similarly, peer-to-peer cortical interactions would require a more complex brain design than the hub-and-spoke architecture.

One might ask: "Is that a problem? Surely there have been half a billion years to evolve the best designs, and in that timescale evolution could have found a complex and efficient neural design.".

However, this argument does not hold. Evolution is a race, and the race goes to the species which can adapt the fastest. Consider two species which start out respectively with a wave storage of spatial positions, and with a pure neural storage - which may both initially be imprecise, inefficient and so on. Both species then evolve, and their evolution explores the two different design spaces for brains – wave storage of positions, and neural storage. As a first approximation, evolution can add useful design information (i.e. it can refine the designs of both brains) at an approximately equal rate (which is probably of the order of a fraction of a bit of useful design information per generation [Worden 1995]). Both species get better at spatial cognition, as their brain designs improve.

For the species with wave storage of positions, the brain design is simpler. So the space of brain designs, which the evolution of that species explores, is a simpler space than the design space for the pure-neural species. The design space has lower dimensionality, and searching it is easier. Evolution can find the good regions in that design space, faster than it can find the good regions in a larger design space. Even if the pure neural species might eventually evolve to have a more efficient brain design than the species with wave storage, it will never get the chance to do so. Long before it does, the wave storage species will have out-competed it in every ecological niche (as the challenge of spatial cognition is similar for all niches – subject to the same laws of geometry) and it will have become extinct.

Put another way, evolution is a process of local hill-climbing in the space of possible brains, by random steps. If the design space of brains is smaller, there is a higher chance that any random step in brain design will increase fitness. For a more complex design, most random steps are bad steps, decreasing fitness. So a species with a simple brain design can hill-climb faster, and over generations it will out-compete a species with a more complex brain design.

So there is a strong selection pressure to have the simplest brain design possible, for given level of performance – to be agile to win the race of cognitive evolution. Wave storage of 3D positions, with a hub and spoke architecture, is simpler design than pure neural storage – so we would expect it to be strongly favoured in the race.

## 7. Connectivity of the Thalamus

Previous sections have described how in computational terms, for both perception and action, a hub-and-spoke wave storage architecture is both simpler and more effective than a pure neural architecture – so will be doubly favoured by evolution.

I now turn from theoretical computational aspects of the wave storage hypothesis, to the empirical evidence for it.

In the theory of this paper, the main role of the thalamus is the storage of 3D spatial information – and as an 'aggregator' it plays a central role in finding the Bayesian best fit to the location of any object around the animal, using sense data of different modalities.

Which modalities? Sense data of most modalities (including vision, touch, proprioception, and hearing) can place precise constraints on the location of any object. But olfaction generally does not. A smell can come from any direction or distance, and there may be no breeze to define what direction it comes from. Smell is of little use in locating objects precisely.

It is therefore significant that across many species, sense data of all modalities pass through the thalamus – except smell, which goes through the olfactory bulb [Sherman & Guillery 2006]. One could plausibly argue that the thalamus is the gateway to cortex for only those sense data which are used to constrain the positions of things – in this theory, because the main function of the thalamus is the precise representation of locations.

There appears to be no other theory of thalamic function which accounts simply for the fact that olfaction does not pass through the thalamus.

## 8. Shape of the Thalamus and TRN

Many parts of the brain, such as the hippocampus or the cortex, have shapes which are very variable across species. All that appears to matter is their synaptic connectivity, and that connectivity is accomplished in whatever way is most economical of axon length and energy consumption. For instance, in the mouse, the cortex is not folded, but cortical regions are still analogous to the folded primate cortex.

In contrast to this, the shapes of the thalamus and the TRN are conserved across species. Across a wide range of species [Jones 1985] which includes mammals, birds and reptiles, and may include more species:

- The thalamus has approximately equal extension in all three dimensions – it is approximately spherical.
- The TRN has the form of a thin shell surrounding the dorsal thalamus.

These facts fit well with the wave hypothesis. If the thalamus holds a wave excitation representing positions, then it needs to have approximately equal extension in all three directions, to accommodate an equal range of wave vectors (i.e. equal spatial precision) in all three directions. If the TRN is the transmitter of the wave, then it needs to be a thin shell surrounding the thalamus, to immerse the thalamus in the wave.

The wave theory makes sense of the distinctive and well conserved anatomy of the thalamus and TRN. As will be described in the next section, a purely neural model of thalamic function does not.

## 9. The Exploding Thalamus

If all the functionality of the thalamus were achieved by neural synaptic connections, then in terms of brain energy consumption, the anatomy of the thalamus and the TRN would not make sense.

Over evolutionary time and in morphogenesis, parts of the brain arrange themselves so as to minimise net axon length between them, to minimise the energy costs of the axons, subject to the required neural connectivity [Van Essen 1997, Attwell & Laughlin 2001, Chklovskii et al. 2002]. This accounts for many aspects of brain anatomy and physiology, such as the folding of the cortex in larger mammals.

From this standpoint of brain energetics, the anatomy of the thalamus and the TRN do not make sense.

The thalamus consists of a number of different primary and secondary nuclei, with no direct connections between the nuclei [Sherman & Guillery 2006] Why then are the nuclei clustered together? Aggregate axon length could be reduced (and so brain energy costs could be reduced) if, over evolutionary time, the thalamic nuclei had migrated out towards cortex (exploded), reducing the net length of thalamo-cortical and cortico-thalamic axons.

Similarly, the TRN consists of a thin shell around the dorsal thalamus, with no neurons projecting from the TRN outwards to cortex [Jones 1987]. All TRN neurons project inwards to the thalamus. Axonal energy costs would be reduced if the TRN was not entirely outside the thalamus, but had migrated some distance towards its centre.

These considerations have been confirmed by approximate numerical calculations of aggregate axon lengths in *homo sapiens* in [Worden 2010, 2014]. Exploding the thalamus would reduce axonal energy costs. They remain to be tested by more precise computations using connectome data.

The distinctive anatomy of the thalamus and the TRN make good sense if the thalamus is immersed in a wave, with the TRN as transmitter. In that case, the thalamic nuclei need to stay close together, and to be inside the TRN, in order to be all immersed in the same wave. This further supports the hypothesis that the thalamus holds a wave, and is reinforced by the wide cross-species preservation of the shape of the thalamus.

## 10. The Form of Consciousness

This section is a brief summary of a related paper [Worden 2019], which may be read in conjunction with this paper – because the form of consciousness we experience may be one of the strongest pieces of evidence for a wave excitation in the thalamus.

There is one very important observation about the form of consciousness, which is rarely noted, but is central to our lived experience. A large part of our conscious experience is **just like the real world**. Consciousness is a model of current reality; it is not just a random set of qualia, like a Jackson Pollock painting.

An important part of our consciousness consists of 3-dimensional geometric model of the world around us [Rudrauf et al 2017, Williford et al 2018] – and it seems to be a rather precise and faithful model.

Most current theories of consciousness , such as [Baars 1988, Tononi 2012], which are based on the classical dynamics of neurons in the brain, do not attempt to account for this important experimental fact, and would have great difficulty in doing so.

However, if there is a wave excitation in the thalamus – whose purpose is to hold a faithful 3-D model of local space - and if consciousness was caused by this wave excitation, then one would expect the form of consciousness to be a faithful 3-D model of local space, as we actually experience it.

Linking consciousness to a wave excitation in the thalamus would give natural and unforced agreement with the key fact about the form of consciousness – much better agreement than other current theories. This makes it a promising line of enquiry, and it is explored further in [Worden 2019].

## 11. Conclusions

Evolution has resulted in a series of elegant, appropriate and often simple designs, from the most basic and universal features of life (such as the form of DNA) onwards.

The core task of any animal brain is to understand where things are in local space, to control the animal's movements. A wave excitation can provide an accurate, simple and computable representation of the locations of things in 3D space.

Compared with a wave representation, a purely neural representation of 3D space would be complicated and inefficient. For the reasons described in this paper, we should consider the possibility that the brain represents local 3D space directly and simply - by a wave excitation in the thalamus.

## Acknowledgements

I should like to thank Karl Friston for very helpful comments and suggestions on previous drafts.